\documentclass[prd,aps,twocolumn,a4paper,floatfix,showpacs,nofootinbib]{revtex4-1}

\usepackage[utf8]{inputenc}  
\usepackage[T1]{fontenc}
\usepackage{graphicx,psfrag}
\usepackage{mathrsfs}
\usepackage{amsmath,amsfonts,amssymb,gensymb}
\usepackage{multirow,enumerate}
\usepackage{comment,hyperref}
\usepackage{color}
\usepackage{xcolor}
\usepackage{acronym}
\usepackage{xspace}
\usepackage[normalem]{ulem}
\usepackage{mathtools}
\usepackage{subfigure}
\usepackage{makecell}
\usepackage{appendix}
\usepackage{array}

\newacro{BH}{black hole}
\newacro{NS}{neutron star}
\newacro{PN}{Post-Newtonian}
\newacro{BBH}{binary black hole}
\newacro{BNS}{binary neutron star}
\newacro{EOB}{effective-one-body}
\newacro{NR}{numerical relativity}
\newacro{GW}{gravitational wave}
\newacro{EOS}{equation-of-state}

\newcommand{\be}{\begin{equation}}
\newcommand{\ee}{\end{equation}}
\newcommand{\bea}{\begin{eqnarray}}
\newcommand{\eea}{\end{eqnarray}}
\newcommand{\bel}{\begin{align}}
\newcommand{\eel}{\end{align}}

\mathchardef\mhyphen="2D

\def\i{{\rm i}}

\def\GMc2{{\rm G M_{\odot} c^{-2}}}

\def\seobNRv4T{\texttt{SEOBNRv4T}\xspace}

\DeclareRobustCommand{\N}[1]{\IfEqCase{#1}{{GW150914}{GW150914\xspace}{GW190620}{GW190620\_030421\xspace}{GW191222}{GW191222\_033537\xspace}{GW190519}{GW190519\_153544\xspace}{GW190521a}{GW190521\xspace}{GW190521b}{GW190521\_074359\xspace}{GW190620}{GW190620\_030421\xspace}{GW190630}{GW190630\_185205\xspace}{GW190910}{GW190910\_112807\xspace}{GW191109}{GW191109\_010717\xspace}{GW191222}{GW191222\_033537\xspace}{GW200112}{GW200112\_155838\xspace}{GW200129}{GW200129\_065458\xspace}{GW200224}{GW200224\_222234\xspace}{GW200311}{GW200311\_115853\xspace}}}

\def\seob{\texttt{SEOBNRv4PHM}\xspace}
\def\nrsur{\texttt{NRSur7dq4}\xspace}
\def\tphm{\texttt{IMRPhenomTPHM}\xspace}
\def\xphm{\texttt{IMRPhenomXPHM}\xspace}

\usepackage{color}
\definecolor{cyan}{rgb}{0,0.9,0.9}
\definecolor{orange}{rgb}{0.9,0.5,0}
\definecolor{magenta}{rgb}{1,0,1}
\definecolor{purple}{rgb}{0.8,0.4,0.8}
\definecolor{gray}{rgb}{0.5,0.5,0.5}
\definecolor{mygreen}{rgb}{0.1,0.8,0.1}
\definecolor{darkblue}{rgb}{0.0,0.0,0.6}

\begin{document}

\title{Comparing gravitational waveform models for binary black hole mergers through a hypermodels approach} 

\author{Anna Puecher$^{1,2}$}
\author{Anuradha Samajdar$^{2,3}$}
\author{Gregory Ashton$^{4}$}
\author{Chris Van Den Broeck$^{1,2}$}
\author{Tim Dietrich$^{3,5}$}

\affiliation{${}^1$Nikhef -- National Institute for Subatomic Physics, 
	Science Park 105, 1098 XG Amsterdam, Netherlands}
\affiliation{${}^2$Institute for Gravitational and Subatomic Physics (GRASP), 
	Utrecht University, Princetonplein 1, 3584 CC Utrecht, Netherlands}
\affiliation{${}^3$Institut f\"{u}r Physik und Astronomie, Universit\"{a}t Potsdam, Haus 28, Karl-Liebknecht-Str. 24/25, 14476, Potsdam, Germany}
\affiliation{${}^4$Royal Holloway, University of London, London TW20 0EX, United Kingdom}
\affiliation{${}^5$Max Planck Institute for Gravitational Physics (Albert Einstein Institute), Am M\"uhlenberg 1, Potsdam 14476, Germany}

\date{\today}

\begin{abstract}
	
The inference of source parameters from gravitational-wave signals relies on theoretical models that describe the emitted waveform. Different model assumptions on which the computation of these models is based could lead to biases in the analysis of gravitational-wave data.
In this work, we sample directly on four state-of-the-art binary black hole waveform models from different families, in order to investigate these systematic biases from the 13 heaviest gravitational-wave sources with moderate to high signal-to-noise ratios in the third Gravitational-Wave Transient Catalog (GWTC-3). All models include spin-precession as well as higher-order modes. Using the ``hypermodels'' technique, we treat the waveform models as one of the sampled parameters, therefore directly getting the odds ratio of one waveform model over another from a single parameter estimation run. From the joint odds ratio over all 13 sources, we find the model \nrsur to be favoured over \seob, with an odds ratio of 29.43; \xphm and \tphm have an odds ratio, respectively, of 4.70 and 5.09 over \seob. However, this result is mainly determined by three events that show a strong preference for some of the models and that are all affected by possible data quality issues. If we do not consider these potentially problematic events, the odds ratio do not exhibit a significant preference for any of the models. We also highlight that the models are not used at their full capabilities since, in order to compare them, we consider only the subdominant modes present in all of them. Although further work studying a larger set of signals will be needed for robust quantitative results, the presented method highlights one possible avenue for future waveform model development.


\end{abstract}

\maketitle

\section{Introduction}
Following the direct detection of gravitational waves (GWs) in 2015~\cite{LIGOScientific:2016aoc, LIGOScientific:2016dsl}, we have direct access and constraints on strong-field gravity~\cite{LIGOScientific:2016lio}. 

With close to 90 significant observations of binary black hole (BBH) mergers~\cite{LIGOScientific:2021djp}, hyperparameters characterizing population models~\cite{LIGOScientific:2021psn} as well as more stringent bounds on strong-field gravity parameters from combining multiple events~\cite{LIGOScientific:2021sio} have been estimated. Ongoing and future observing runs of the LIGO-Scientific, Virgo, and KAGRA collaborations will operate at higher sensitivities and enable us to see many more events. However, as the statistical biases reduce through improved detector sensitivities and by combining multiple events, the systematic effects from the GW models employed to analyze our data will start dominating. Several studies have been made to expose this problem with future-generation detectors, e.g.~Ref.~\cite{Purrer:2019jcp}. 

Typically, GW source properties are inferred by analyzing the data with multiple waveform models where the estimates broadly agree. This serves as a consistency test between different models developed employing different techniques. Separate analyses are therefore performed on a single event to obtain estimates of the same. However, while individual sources may be consistent, combining the data may expose a bias or preference for one model over another. 
In this work, we infer the parameter properties of the 13 heaviest and significant BBH observations by Advanced LIGO~\cite{LIGOScientific:2014pky} and Advanced Virgo~\cite{VIRGO:2014yos} in GWTC-3~\cite{LIGOScientific:2021djp} and quantify the preference for one waveform model over another from the combined GW data. The choice of events is determined by the fact that, for one of the models employed, the region of validity covers only high values of the binary's total mass; moreover, the shorter duration of signals produced by high-mass systems reduces the computational cost of the analysis.
Reference~\cite{Hoy:2022tst} has looked at a very similar problem from a technical point of view, performing a joint Bayesian analysis with three different models on a large set of simulated events, showing consistent results with the ones obtained via a Bayesian model averaging method, and with a significant gain in terms of computational cost.  However, the analyzed signals were all simulations apart from one real GW event, \N{GW200129}, also included in our suite of events. 
The focus of our work is instead on real events, with the goal to investigate possible systematic biases caused by the different waveform approximants. We employ four waveform models: \nrsur~\cite{Varma:2019csw}, \xphm~\cite{Pratten:2020ceb}, \tphm~\cite{Estelles:2021gvs}, and \seob~\cite{Ossokine:2020kjp}. 
In Ref.~\cite{Islam:2023zzj}, all the events in GWTC-3 are analyzed with the \nrsur model, finding, in some cases, different results with respect to the ones obtained with the \xphm and \seob models in the LVK analyses.

For our study, we focus on the method introduced in Ref.~\cite{Ashton:2021cub}, henceforth referred to as \emph{hypermodels}. The purpose of our study is to obtain a quantitative measure of selection, in this case by using the \emph{odds ratio}, between one waveform and another from a combination of GW events.   

We outline our analysis method in Sec.~\ref{sec:methods}, by giving an overview of the models used in Sec.~\ref{ssec:wfrms} and the inference techniques in Sec.~\ref{ssec:pe}. We summarize our results on the individual events and the combined analyses in Sec.~\ref{sec:res}. We conclude in Sec.~\ref{sec:summary}. In Appendix~\ref{sec:appendix} we show results of injection runs in order to validate our method.

\section{Methods}
\label{sec:methods}

\subsection{Waveform models}
\label{ssec:wfrms}

We consider for our analysis four state-of-the-art BBH waveforms, all including precession~\cite{Hannam:2013oca,Schmidt:2014iyl} and higher-order modes~\cite{Blanchet:2008je}. The construction of the precessing approximant is usually based on a non-precessing one. The specific subdominant modes $(\ell,|m|)$ included, and listed below, are the ones provided by the aligned-spin model: when constructing the precessing waveform, it will include all the higher-order modes corresponding to a given $\ell$, although their description might be incomplete based on the mode content of the aligned-spin approximant.

We note that waveform models do not strictly have identical definitions for the underlying parameters. As such, within the hypermodel approach (and indeed any waveform systematic study), care should be taken when comparing posterior inferences.

The employed models are briefly described below.

\subsubsection{NRSur7dq4}

\nrsur \cite{Varma:2019csw} is a time-domain surrogate model that extends the previous \texttt{NRSur7dq2}~\cite{Blackman:2017pcm} to higher values of mass ratio. Surrogate models \cite{Field:2013cfa,Blackman:2015pia} are constructed by interpolating over a set of precomputed waveforms, in this case numerical-relativity (NR) waveforms built over the parameter space for precessing BBH systems. This approach produces very accurate waveforms, since it does not rely on any approximation, except for the numerical discretization in the simulations. However, due to the computational cost of NR simulations, only a limited parameter space region can be covered. In particular, the \nrsur model is valid for mass ratio values up to $q \le 6$ and for total mass values $M \gtrsim 66 \, M_\odot$ (cf.~Fig.~9 in Ref.\cite{Varma:2019csw} for the precise range of validity as a function of the system's mass ratio). \nrsur includes, in the co-precessing frame, all the subdominant modes up to $\ell \le 4$.

\subsubsection{SEOBNRv4PHM}

\seob \cite{Ossokine:2020kjp} is a time-domain, effective-one-body precessing waveform built from the aligned-spin model in Ref.~\cite{Cotesta:2018fcv}. The effective-one-body formalism (EOB) \cite{Pan:2013rra,Babak:2016tgq} maps the dynamics of two bodies into the dynamic of a reduced-mass body moving in a deformed metric. The gravitational waveforms computed with this approach are accurate but slow to generate.
For \seob, the precessing sector is not calibrated to NR simulations.
In the co-precessing frame, it includes the subdominant harmonics $(\ell,|m|) = (2,1), (3,3), (4,4), (5,5)$, and it is valid for mass ratio values in the range $1 \le q \le 50$.

\subsubsection{IMRPhenomXPHM}

\xphm \cite{Pratten:2020ceb} is a phenomenological, frequency-domain approximant based on the non-precessing \texttt{IMRPhenomXHM} model \cite{Garcia-Quiros:2020qpx}, and constructed via the so-called ``twisting-up'' procedure \cite{Hannam:2013oca,Schmidt:2012rh,Schmidt:2010it}, which allows to map non-precessing systems to precessing ones.
Phenomenological models \cite{Ajith:2009bn,Santamaria:2010yb} are built from piecewise closed-form expressions, which make them computationally cheap. \texttt{IMRPhenomXHM} is constructed separately for the three different inspiral, intermediate, and ringdown regions. The intermediate region is fully calibrated to NR simulations, while the inspiral and ringdown ones also include information from the post-Newtonian expansion or black hole perturbation theory, respectively.
In the co-precessing frame, this approximant includes the subdominant modes $(\ell,|m|) = (2,1),(3,3),(3,2),(4,4)$, which are calibrated to NR waveforms individually. The model is valid for spins magnitude up to 0.99 and $q \le 1000$ (while its recommended usage region is $q \le 20$, due to its calibration to NR simulations).

\subsubsection{IMRPhenomTPHM}

This approximant also belongs to the family of phenomenological models, but it is built in the time domain. Although working in the frequency domain offers an additional speed-up when computing the noise-weighted inner products, a time-domain model allows a direct description of the system's dynamics. \tphm \cite{Estelles:2021gvs} is built from the non-precessing model \texttt{IMRPhenomTHM} \cite{Estelles:2020twz} via the ``twisting-up'' procedure, which is however different to the procedure applied in the frequency domain. In the co-precessing frame, this model includes the subdominant harmonics $(\ell, |m|) = (2,1),(3,3),(4,4),(5,5)$. The parameter range of validity is defined by: $m_2 \ge 0.5 \, M_\odot$, with $m_2$ being the secondary mass, and spin magnitude $|\chi_{1,2}| \le 0.99$ for $q \le 200$ (while its recommended usage region is $q \le 20$, due to its calibration to NR simulations).

\subsection{Bayesian framework}
\label{ssec:pe}

Analyzing GW signals in a Bayesian framework allows both inference of the source parameters and a comparison between different possible models describing the GW waveform. The source parameters $\vec{\theta}$ can be recovered from the detector data $d$ evaluating the posterior $p(\vec{\theta}|d, \Omega)$, where $\Omega$ is the waveform model. In this context, Bayes theorem reads
\begin{equation}
	p(\vec{\theta}|d, \Omega) = \frac{p(d | \vec{\theta}, \Omega) p(\vec{\theta}|\Omega)}{p(d | \Omega )},
	\label{eq:bayes}
\end{equation}
where $p(d | \vec{\theta}, \Omega)$ represents the \emph{likelihood} of observing the data $d$ given the model $\Omega$ and the specific set of parameters $\vec{\theta}$, and $p(\vec{\theta}|\Omega)$ the \emph{prior probability density}. We employed the same default priors used in the parameter estimation analysis for these events in the LVK catalog papers~\cite{LIGOScientific:2021usb,LIGOScientific:2021djp}, adjusting them as follows in order to respect the region of validity of all the four approximants considered: $q \le 6$,  $\chi_{1,2} \le 0.99$, $m_2 \ge 0.5 M_\odot$. For some events, we also adjust the prior on chirp mass to ensure $\mathcal{M}_c \ge 26 M_\odot$, to allow for the validity of \nrsur in the entire region of the prior volume.
The denominator in Eq.~\ref{eq:bayes} is the \emph{evidence} for the model $\Omega$, and is determined by the requirement that the posterior distribution must be normalized
\begin{equation}
	p(d | \Omega ) = \int d\vec{\theta} \, p(d | \vec{\theta}, \Omega) p(\vec{\theta}|\Omega).
\end{equation}
The evidence allows us to compare different models, say $\Omega_A$ and $\Omega_B$, computing the \emph{odds ratio}
\begin{equation}
	\mathcal{O}^A_B = \frac{p(\Omega_A | d)}{p(\Omega_B | d)} = \underbrace{\frac{p(d | \Omega_A)}{p(d | \Omega_B)}}_{\mathcal{B}_B^A} \underbrace{\frac{p(\Omega_A)}{p(\Omega_B)}}_{\pi_B^A} = \mathcal{B}_B^A \times \pi_B^A,
\end{equation}
where the \emph{Bayes factor} $\mathcal{B}_B^A$ is the ratio of the evidence for the two models given the data, and $\pi_B^A$ is usually set to 1, meaning that we do not have any a-priori preference for one of the models.

The posterior probability density and the evidence can be estimated with stochastic sampling methods. In particular, here we employ the \emph{hypermodels} approach introduced in Ref.~\cite{Ashton:2021cub}, with a Metropolis-Hastings MCMC algorithm~\cite{Metropolis:1953am,Hastings:1970aa}, based on the implementation of the \textsc{Bilby}-MCMC sampler~\cite{Ashton:2021anp}. 

\subsection{Hypermodels}

The waveform model $\Omega$ employed during the sampling is substituted with a hypermodel $\Omega = \left\{\Omega_0, \Omega_1, ..., \Omega_{n-1}\right\}$, with $n$ being the number of models we want to study. The parameter space investigated by the sampler, therefore, becomes $\{\vec{\theta},\omega\}$, where $\vec{\theta}$ are the usual source parameters, while $\omega$ is a categorical parameter $\omega \in [0,1,...,n-1]$ representing the waveform approximant. We define a mapping between the value of the parameter $\omega$ and a specific waveform approximant, so that at each iteration the sampler picks a value of $\{\vec{\theta},\omega\}$ and generates the waveform with parameters $\vec{\theta}$ and the approximant corresponding to $\omega$. We employ an uninformative prior $\pi(\omega) = 1/n$, which translates into a prior odds $\pi_B^A = 1$ for all the combinations of models considered. Among the final $N$ posterior samples, we can distinguish the samples for each waveform $\ell$ from the value of the $\omega$ parameter. If $n_\ell$ is the number of samples for the $\ell\mhyphen$th approximant, its probability with respect to the other waveforms is given by $p_\ell = n_\ell/N$.
The odds ratio between two models $\omega = A$ and $\omega = B$ is computed as
\begin{equation}
	\mathcal{O}_B^A = \frac{p_A}{p_B} = \frac{n_A}{n_B},
\end{equation}

The error on $p_{A,B}$ is given by the variance of the mean of a Poisson process, yielding $\sigma^2_{p_A,p_B} = p_{A,B}/N$. For two random variables $v_1$ and $v_1$, with a standard deviation $\sigma_1$ and $\sigma_2$, respectively, one can compute the the standard deviation on their ratio as
	
	\begin{equation}
		\sigma^2 _{\frac{v_1}{v_2}} = \frac{v_1}{v_2} \left[ \left(\frac{\sigma_1}{v_1}\right)^2 + \left(\frac{\sigma_2}{v_2}\right)^2 - 2\frac{\sigma_{12}}{v_1 v_2}\right],
	\end{equation}
where $\sigma_{12}$ is the covariance.
Therefore, propagating the uncertainty, and ignoring any correlation, the variance for the odds ratio $\mathcal{O}_B^A$ is given by 
\begin{align}
	\sigma^2_{\mathcal{O}_B^A} = \sigma^2_{\frac{p_A}{p_B}} &\approx \left( \frac{p_A}{p_B}\right)^2 \left[ \left(\frac{\sigma_{p_A}}{p_A}\right)^2 + \left(\frac{\sigma_{p_B}}{p_B}\right)^2 \right] \\
	&\approx \mathcal({O}_B^A)^2 \left( \frac{p_A}{N} \frac{1}{p_A^2} + \frac{p_B}{N} \frac{1}{p_B^2}\right) \\
	&\approx \frac{(\mathcal{O}_B^A)^2}{N}\left(\frac{1}{p_A} + \frac{1}{p_B} \right).
\end{align}


\begingroup
\renewcommand*{\arraystretch}{2}
\begin{table*}
\begin{tabular}{c|c c c c c|c c c c c }
\hline
&\multicolumn{5}{c|}{\centering{$\mathcal{M}_c \, [M_\odot]$}} & \multicolumn{5}{ c }{\centering{$q$}} \\ \hline
  Event & NRSur & SEOB & IMRX & IMRT & Combined & NRSur & SEOB & IMRX & IMRT & Combined \\ \hline
\N{GW150914} & $31.0^{+1.1}_{-1.2}$ & $30.6^{+1.6}_{-1.5}$ & $30.6^{+1.3}_{-1.6}$ & $31.1^{+1.2}_{-1.2}$ & $30.9^{+1.3}_{-1.5}$ & $0.9^{+0.1}_{-0.2}$ & $0.9^{+0.1}_{-0.2}$ & $0.9^{+0.1}_{-0.2}$ & $0.9^{+0.1}_{-0.2}$ & $0.9^{+0.1}_{-0.2}$ \\
\N{GW190519} & $66.4^{+6.7}_{-11.6}$ & $66.2^{+8.1}_{-12.0}$ & $64.6^{+7.8}_{-10.6}$ & $67.5^{+7.4}_{-12.5}$ & $65.7^{+7.8}_{-11.4}$ & $0.6^{+0.3}_{-0.2}$ & $0.6^{+0.2}_{-0.2}$ & $0.6^{+0.3}_{-0.2}$ & $0.6^{+0.2}_{-0.2}$ & $0.6^{+0.3}_{-0.2}$ \\
\N{GW190521b} & $40.4^{+2.0}_{-3.2}$ & $40.7^{+2.8}_{-2.7}$ & $39.4^{+2.3}_{-2.4}$ & $40.8^{+1.9}_{-3.0}$ & $40.4^{+2.5}_{-2.9}$ & $0.8^{+0.2}_{-0.2}$ & $0.8^{+0.2}_{-0.2}$ & $0.8^{+0.2}_{-0.2}$ & $0.8^{+0.2}_{-0.2}$ & $0.8^{+0.2}_{-0.2}$ \\
\N{GW190620} & $58.6^{+7.2}_{-10.9}$ & $60.3^{+9.8}_{-10.3}$ & $58.9^{+9.2}_{-12.9}$ & $60.1^{+6.6}_{-11.0}$ & $59.5^{+8.2}_{-11.3}$ & $0.7^{+0.3}_{-0.3}$ & $0.7^{+0.3}_{-0.3}$ & $0.6^{+0.3}_{-0.3}$ & $0.7^{+0.3}_{-0.3}$ & $0.7^{+0.3}_{-0.3}$ \\
\N{GW190630} & $29.5^{+1.5}_{-1.8}$ & $29.3^{+1.9}_{-1.9}$ & $29.4^{+1.7}_{-1.6}$ & $29.6^{+1.6}_{-1.8}$ & $29.5^{+1.6}_{-1.8}$ & $0.7^{+0.3}_{-0.2}$ & $0.6^{+0.3}_{-0.2}$ & $0.7^{+0.3}_{-0.2}$ & $0.7^{+0.3}_{-0.2}$ & $0.7^{+0.3}_{-0.2}$ \\
\N{GW190910} & $43.3^{+3.6}_{-3.7}$ & $43.3^{+3.9}_{-3.7}$ & $43.2^{+4.1}_{-4.2}$ & $43.5^{+3.9}_{-3.5}$ & $43.3^{+3.9}_{-3.8}$ & $0.8^{+0.2}_{-0.2}$ & $0.8^{+0.2}_{-0.2}$ & $0.8^{+0.2}_{-0.2}$ & $0.8^{+0.2}_{-0.2}$ & $0.8^{+0.2}_{-0.2}$ \\
\N{GW191222} & $52.6^{+5.4}_{-6.2}$ & $51.6^{+7.3}_{-6.6}$ & $51.0^{+6.6}_{-7.0}$ & $52.8^{+5.6}_{-5.9}$ & $52.2^{+6.1}_{-6.6}$ & $0.8^{+0.2}_{-0.3}$ & $0.8^{+0.2}_{-0.3}$ & $0.8^{+0.2}_{-0.3}$ & $0.8^{+0.2}_{-0.3}$ & $0.8^{+0.2}_{-0.3}$ \\
\N{GW200112} & $33.8^{+2.5}_{-1.9}$ & $34.1^{+3.4}_{-2.5}$ & $33.8^{+2.6}_{-2.3}$ & $34.0^{+2.7}_{-2.0}$ & $33.9^{+2.8}_{-2.1}$ & $0.8^{+0.2}_{-0.2}$ & $0.8^{+0.2}_{-0.3}$ & $0.8^{+0.2}_{-0.3}$ & $0.8^{+0.2}_{-0.2}$ & $0.8^{+0.2}_{-0.2}$ \\
\N{GW200224} & $40.3^{+3.9}_{-3.8}$ & $40.7^{+3.7}_{-3.8}$ & $40.6^{+3.1}_{-3.7}$ & $40.3^{+4.4}_{-3.8}$ & $40.5^{+3.6}_{-3.8}$ & $0.8^{+0.2}_{-0.3}$ & $0.8^{+0.2}_{-0.2}$ & $0.8^{+0.2}_{-0.2}$ & $0.8^{+0.2}_{-0.3}$ & $0.8^{+0.2}_{-0.3}$ \\
\N{GW200311} & $32.7^{+2.6}_{-2.9}$ & $32.6^{+2.8}_{-2.6}$ & $32.4^{+2.6}_{-2.7}$ & $33.1^{+2.9}_{-3.2}$ & $32.6^{+2.8}_{-2.8}$ & $0.8^{+0.2}_{-0.3}$ & $0.8^{+0.2}_{-0.3}$ & $0.8^{+0.2}_{-0.3}$ & $0.8^{+0.2}_{-0.3}$ & $0.8^{+0.2}_{-0.3}$ \\ 
\N{GW190521a} & $112.8^{+12.1}_{-13.2}$ & $119.3^{+18.9}_{-16.9}$ & $104.5^{+16.9}_{-14.4}$ & $114.5^{+18.7}_{-14.8}$ & $114.5^{+18.9}_{-15.4}$ & $0.8^{+0.1}_{-0.3}$ & $0.7^{+0.2}_{-0.2}$ & $0.7^{+0.3}_{-0.1}$ & $0.8^{+0.2}_{-0.2}$ & $0.8^{+0.2}_{-0.2}$ \\
\N{GW191109} & $60.3^{+5.6}_{-9.4}$ & $62.2^{+9.1}_{-7.5}$ & $59.4^{+13.5}_{-8.4}$ & $66.3^{+6.8}_{-8.4}$ & $62.9^{+9.0}_{-8.2}$ & $0.7^{+0.2}_{-0.3}$ & $0.7^{+0.2}_{-0.2}$ & $0.8^{+0.2}_{-0.2}$ & $0.8^{+0.2}_{-0.2}$ & $0.7^{+0.2}_{-0.3}$ \\
\N{GW200129} & $29.9^{+2.5}_{-1.5}$ & $31.6^{+0.8}_{-1.3}$ & $31.7^{+2.3}_{-3.1}$ & $31.4^{+1.8}_{-1.8}$ & $30.9^{+2.8}_{-2.4}$ & $0.5^{+0.4}_{-0.1}$ & $0.8^{+0.2}_{-0.4}$ & $0.7^{+0.3}_{-0.3}$ & $0.8^{+0.1}_{-0.2}$ & $0.6^{+0.4}_{-0.2}$ \\\hline
 
\end{tabular}
\caption{Median values and their 5\% and 95\% quantiles from the probability density functions of mass parameters, chirp mass $\mathcal{M}_c$ and mass ratio $q$, for the different models' posteriors and for the combined one.}
\label{tab:mass-pars}
\end{table*}
\endgroup

\begingroup
\renewcommand*{\arraystretch}{2}
\begin{table*}
\begin{tabular}{c|c c c c c|c c c c c }
\hline
&\multicolumn{5}{c|}{\centering{$\chi_{\rm eff}$}} & \multicolumn{5}{c}{\centering{$\chi_{\rm p}$}} \\ \hline
  Event & NRSur & SEOB & IMRX & IMRT & Combined & NRSur & SEOB & IMRX & IMRT & Combined \\ \hline
\N{GW150914} & $-0.02^{+0.09}_{-0.11}$ & $-0.03^{+0.11}_{-0.12}$ & $-0.04^{+0.10}_{-0.14}$ & $-0.01^{+0.09}_{-0.10}$ & $-0.02^{+0.10}_{-0.12}$ & $0.35^{+0.44}_{-0.27}$ & $0.33^{+0.43}_{-0.25}$ & $0.50^{+0.39}_{-0.39}$ & $0.39^{+0.42}_{-0.31}$ & $0.39^{+0.44}_{-0.31}$ \\
\N{GW190519} & $0.31^{+0.20}_{-0.23}$ & $0.34^{+0.21}_{-0.26}$ & $0.33^{+0.19}_{-0.26}$ & $0.31^{+0.21}_{-0.26}$ & $0.33^{+0.20}_{-0.25}$ & $0.50^{+0.33}_{-0.32}$ & $0.45^{+0.35}_{-0.27}$ & $0.47^{+0.36}_{-0.29}$ & $0.52^{+0.34}_{-0.34}$ & $0.48^{+0.35}_{-0.30}$ \\
\N{GW190521b} & $0.12^{+0.11}_{-0.13}$ & $0.15^{+0.11}_{-0.12}$ & $0.08^{+0.12}_{-0.11}$ & $0.16^{+0.10}_{-0.14}$ & $0.14^{+0.11}_{-0.14}$ & $0.44^{+0.34}_{-0.31}$ & $0.43^{+0.37}_{-0.29}$ & $0.32^{+0.39}_{-0.25}$ & $0.45^{+0.36}_{-0.31}$ & $0.42^{+0.37}_{-0.30}$ \\
\N{GW190620} & $0.32^{+0.22}_{-0.25}$ & $0.39^{+0.20}_{-0.22}$ & $0.35^{+0.20}_{-0.28}$ & $0.37^{+0.19}_{-0.23}$ & $0.35^{+0.21}_{-0.25}$ & $0.51^{+0.35}_{-0.33}$ & $0.46^{+0.35}_{-0.30}$ & $0.54^{+0.35}_{-0.36}$ & $0.46^{+0.33}_{-0.29}$ & $0.49^{+0.35}_{-0.32}$ \\
\N{GW190630} & $0.10^{+0.13}_{-0.14}$ & $0.10^{+0.14}_{-0.14}$ & $0.09^{+0.13}_{-0.13}$ & $0.11^{+0.14}_{-0.15}$ & $0.10^{+0.13}_{-0.14}$ & $0.34^{+0.40}_{-0.25}$ & $0.30^{+0.35}_{-0.22}$ & $0.30^{+0.38}_{-0.23}$ & $0.31^{+0.34}_{-0.23}$ & $0.31^{+0.37}_{-0.23}$ \\
\N{GW190910} & $-0.02^{+0.17}_{-0.18}$ & $0.00^{+0.16}_{-0.20}$ & $-0.01^{+0.17}_{-0.20}$ & $0.00^{+0.18}_{-0.18}$ & $-0.01^{+0.17}_{-0.19}$ & $0.43^{+0.42}_{-0.34}$ & $0.39^{+0.39}_{-0.32}$ & $0.39^{+0.45}_{-0.31}$ & $0.40^{+0.42}_{-0.32}$ & $0.41^{+0.42}_{-0.32}$ \\
\N{GW191222} & $-0.03^{+0.19}_{-0.22}$ & $-0.01^{+0.20}_{-0.25}$ & $-0.05^{+0.19}_{-0.24}$ & $-0.02^{+0.19}_{-0.20}$ & $-0.02^{+0.19}_{-0.23}$ & $0.41^{+0.44}_{-0.32}$ & $0.41^{+0.43}_{-0.32}$ & $0.40^{+0.42}_{-0.30}$ & $0.42^{+0.43}_{-0.33}$ & $0.41^{+0.44}_{-0.32}$ \\
\N{GW200112} & $0.04^{+0.15}_{-0.13}$ & $0.07^{+0.17}_{-0.15}$ & $0.05^{+0.14}_{-0.15}$ & $0.06^{+0.16}_{-0.13}$ & $0.06^{+0.16}_{-0.14}$ & $0.36^{+0.42}_{-0.28}$ & $0.35^{+0.41}_{-0.28}$ & $0.39^{+0.45}_{-0.30}$ & $0.36^{+0.41}_{-0.28}$ & $0.36^{+0.43}_{-0.28}$ \\
\N{GW200224} & $0.09^{+0.17}_{-0.15}$ & $0.11^{+0.14}_{-0.16}$ & $0.10^{+0.14}_{-0.16}$ & $0.11^{+0.17}_{-0.16}$ & $0.10^{+0.15}_{-0.16}$ & $0.43^{+0.41}_{-0.34}$ & $0.39^{+0.42}_{-0.30}$ & $0.48^{+0.39}_{-0.35}$ & $0.38^{+0.41}_{-0.30}$ & $0.44^{+0.41}_{-0.33}$ \\
\N{GW200311} & $-0.02^{+0.16}_{-0.19}$ & $-0.01^{+0.15}_{-0.19}$ & $-0.04^{+0.16}_{-0.19}$ & $0.01^{+0.17}_{-0.21}$ & $-0.02^{+0.16}_{-0.19}$ & $0.44^{+0.39}_{-0.34}$ & $0.40^{+0.43}_{-0.31}$ & $0.49^{+0.39}_{-0.37}$ & $0.48^{+0.40}_{-0.37}$ & $0.46^{+0.41}_{-0.35}$ \\ 
\N{GW190521a} & $-0.14^{+0.35}_{-0.37}$ & $0.07^{+0.32}_{-0.38}$ & $-0.08^{+0.35}_{-0.46}$ & $-0.17^{+0.38}_{-0.31}$ & $-0.10^{+0.39}_{-0.38}$ & $0.75^{+0.20}_{-0.35}$ & $0.71^{+0.24}_{-0.37}$ & $0.49^{+0.35}_{-0.34}$ & $0.76^{+0.19}_{-0.33}$ & $0.73^{+0.22}_{-0.37}$ \\
\N{GW191109} & $-0.42^{+0.29}_{-0.27}$ & $-0.32^{+0.38}_{-0.26}$ & $-0.33^{+0.59}_{-0.33}$ & $-0.24^{+0.25}_{-0.28}$ & $-0.31^{+0.36}_{-0.28}$ & $0.60^{+0.29}_{-0.26}$ & $0.74^{+0.22}_{-0.36}$ & $0.60^{+0.31}_{-0.35}$ & $0.85^{+0.12}_{-0.33}$ & $0.75^{+0.21}_{-0.37}$ \\
\N{GW200129} & $-0.01^{+0.14}_{-0.11}$ & $0.07^{+0.09}_{-0.04}$ & $0.10^{+0.15}_{-0.18}$ & $0.07^{+0.13}_{-0.13}$ & $0.04^{+0.18}_{-0.16}$ & $0.86^{+0.12}_{-0.35}$ & $0.28^{+0.52}_{-0.13}$ & $0.82^{+0.15}_{-0.39}$ & $0.48^{+0.38}_{-0.34}$ & $0.83^{+0.14}_{-0.41}$ \\ \hline
 
\end{tabular}
\caption{Median values and their 5\% and 95\% quantiles from the probability density functions of spin parameters, effective-spin $\chi_{\mathrm{eff}}$ and spin-precessing parameter $\chi_{\mathrm{p}}$, for the different models' posteriors and for the combined one.}
\label{tab:spin-pars}
\end{table*}
\endgroup

\begingroup
\renewcommand*{\arraystretch}{2}
\begin{table*}
	\begin{tabular}{c|c c c c c}
		\hline
		&\multicolumn{5}{c}{\centering{$\log \mathcal{L}$}} \\ \hline
		Event & NRSur & SEOB & IMRX & IMRT & Combined \\ \hline
\N{GW150914} & $322.2^{+2.7}_{-4.3}$ & $321.6^{+2.5}_{-4.1}$ & $322.2^{+2.8}_{-4.0}$ & $\boldsymbol{322.4^{+2.6}_{-4.4}}$ & $322.2^{+2.7}_{-4.3}$ \\
\N{GW190519} & $114.6^{+3.7}_{-4.9}$ & $\boldsymbol{115.4^{+3.7}_{-5.3}}$ & $115.1^{+3.3}_{-5.1}$ & $114.6^{+3.2}_{-5.2}$ & $115.0^{+3.5}_{-5.1}$ \\
\N{GW190521b} & $320.0^{+3.5}_{-4.8}$ & $\boldsymbol{321.3^{+3.2}_{-5.1}}$ & $319.7^{+3.1}_{-4.4}$ & $320.6^{+3.4}_{-4.6}$ & $320.6^{+3.5}_{-4.8}$ \\
\N{GW190620} & $64.1^{+3.9}_{-5.3}$ & $64.0^{+4.1}_{-5.6}$ & $63.7^{+3.6}_{-5.4}$ & $\boldsymbol{64.2^{+3.8}_{-5.6}}$ & $64.0^{+3.9}_{-5.5}$ \\
\N{GW190630} & $\boldsymbol{117.7^{+3.1}_{-5.1}}$ & $116.8^{+3.2}_{-5.1}$ & $116.9^{+3.1}_{-4.9}$ & $117.7^{+3.1}_{-5.0}$ & $117.4^{+3.2}_{-5.1}$ \\
\N{GW190910} & $90.5^{+3.3}_{-4.6}$ & $\boldsymbol{90.8^{+3.9}_{-4.6}}$ & $90.4^{+3.1}_{-4.5}$ & $90.4^{+3.7}_{-4.5}$ & $90.5^{+3.5}_{-4.6}$ \\
\N{GW191222} & $70.0^{+2.5}_{-4.1}$ & $69.5^{+2.5}_{-4.0}$ & $69.3^{+2.5}_{-4.0}$ & $\boldsymbol{70.1^{+2.5}_{-4.1}}$ & $69.8^{+2.6}_{-4.1}$ \\
\N{GW200112} & $166.2^{+2.9}_{-4.4}$ & $165.5^{+2.8}_{-4.6}$ & $165.6^{+2.7}_{-4.4}$ & $\boldsymbol{166.4^{+2.9}_{-4.4}}$ & $166.0^{+2.9}_{-4.5}$ \\
\N{GW200224} & $188.1^{+3.6}_{-4.5}$ & $188.0^{+2.7}_{-4.4}$ & $\boldsymbol{188.6^{+3.3}_{-4.6}}$ & $187.4^{+2.7}_{-4.5}$ & $188.1^{+3.3}_{-4.5}$ \\
\N{GW200311} & $145.4^{+2.7}_{-4.2}$ & $146.0^{+2.6}_{-4.2}$ & $\boldsymbol{146.2^{+2.5}_{-4.3}}$ & $145.6^{+2.8}_{-4.2}$ & $145.9^{+2.7}_{-4.3}$ \\
\N{GW190521a} & $88.0^{+4.2}_{-5.6}$ & $87.4^{+4.2}_{-5.4}$ & $83.6^{+4.3}_{-4.3}$ & $\boldsymbol{88.4^{+3.6}_{-5.5}}$ & $87.8^{+4.1}_{-5.8}$ \\
\N{GW191109} & $133.3^{+3.9}_{-6.2}$ & $\boldsymbol{136.4^{+5.6}_{-6.9}}$ & $132.2^{+6.9}_{-6.6}$ & $135.9^{+5.4}_{-6.7}$ & $135.8^{+5.9}_{-6.9}$ \\
\N{GW200129} & $\boldsymbol{347.2^{+4.4}_{-7.1}}$ & $341.0^{+2.6}_{-3.8}$ & $345.3^{+4.7}_{-6.4}$ & $341.1^{+5.3}_{-4.6}$ & $346.1^{+4.8}_{-7.0}$ \\ \hline
\end{tabular}
\caption{Median values and their 5\% and 95\% quantiles from the probability density functions of the recovered $\log \mathcal{L}$ with the different models and for the combined results. For each event, the highest value of $\log \mathcal{L}$ is marked in bold. }
\label{tab:log_lik}
\end{table*}
\endgroup

\section{Results}
\label{sec:res}

We analyze 13 events of GWTC-3, using the data available on GWOSC~\cite{KAGRA:2023pio,LIGOScientific:2019lzm},focusing on the ones with the highest total mass ($M > 59.4 \, M_\odot$), and with moderate to high signal-to-noise ratios (SNRs). If $h(\vec{\theta})$ is the GW signal, with $\vec{\theta}$ the binary's parameters, the optimal SNR is defined as $\left\langle h(\vec{\theta}) | h(\vec{\theta})\right\rangle ^{1/2}$. In particular, we consider events with a network SNR $\rho_{\rm net} \ge \sqrt{N_{\rm d}  \times 8^2}$, where $N_{\rm d}$ is the number of interferometers detecting the event, corresponding to at least a signal-to-noise ratio 8 per detector. 
The waveform models employed include higher-order modes, and we used the modes available for all models: $(\ell,m)= (2,2), (2,1), (3,3), (4,4), (2,-2), (2,-1), (3,-3), (4,-4)$. We remark that this implies that the models are not used at their full capabilities, since we could not include their full mode content. For \seob, the sampling rate must ensure that the Nyquist frequency is larger than the ringdown frequency. For most events, this means that the required sampling rate was higher than the one used for the LVK catalog papers~\cite{LIGOScientific:2021usb,LIGOScientific:2021djp}; therefore we estimated the events' power spectral densities (PSDs) in the needed frequency range, using \texttt{BayesLine} \cite{Littenberg:2014oda} and the same settings as in Ref.~\cite{LIGOScientific:2021usb,LIGOScientific:2021djp}. The new PSDs are released together with this paper~\cite{puecher_anna_2023_8251823}.

For our analysis runs, the reference frequency is set to 20 Hz, and the waveform templates are generated starting at $f_{\rm low} = 20$~Hz, the same lower frequency used for the analysis, for all the models apart from \seob, for which the waveform needs to be generated starting from lower frequencies (see Sec.IIIC of Ref.~\cite{Ossokine:2020kjp})\footnote{More specifically, following the EOB formalism, the waveform must be generated for an initial radial separation $r > 10.5 \, M$, which, through Kepler's law, translates into a maximum value of the initial frequency for the $(2,2)$ mode~\cite{Ossokine:2020kjp}. Also in the LVK analyses a lower initial frequency for waveform generation is usually employed for the analysis with \seob. Since this condition on the minimum frequency derives directly from the constraint on the initial separation in the EOB formalism, the same does not apply for the other time-domain models. In particular, for \nrsur the region of validity mentioned in Sec.~\ref{ssec:wfrms} holds specifically for a starting frequency of 20 Hz. For what concerns \tphm, instead, the length of the waveform is simply given by the time spent between the starting frequency and the frequency of the amplitude peak of the $(2,2)$ mode, therefore potential issues arise only when the $(2,2)$ peak frequency happens below the specified $f_{\rm low}$.}. 

\begingroup
\renewcommand*{\arraystretch}{2}
\begin{table}
	\begin{tabular}{c|c c c c|c}
		\hline
		Event & NRSur & SEOB & IMRX & IMRT & Combined  \\ \hline
		\N{GW150914} & 0.008 & 0.010 & 0.050 & 0.017 & 0.015  \\ 
		\N{GW190519} & 0.010 & 0.017 & 0.010 & 0.011 & 0.011 \\ 
		\N{GW190521b} & 0.037 & 0.029 & 0.027 & 0.029 & 0.029 \\ 
		\N{GW190620} & 0.010 & 0.006 & 0.016 & 0.012 & 0.006  \\ 
		\N{GW190630} & 0.030 & 0.067 & 0.049 & 0.065 & 0.050  \\ 
		\N{GW190910} & 0.023 & 0.012 & 0.009 & 0.014 & 0.014  \\ 
		\N{GW191222} & 0.012 & 0.011 & 0.014 & 0.012 & 0.011  \\ 
		\N{GW200112} & 0.012 & 0.015 & 0.011 & 0.014 & 0.012 \\ 
		\N{GW200224} & 0.008 & 0.011 & 0.024 & 0.010 & 0.010\\ 
		\N{GW200311} & 0.026 & 0.018 & 0.041 & 0.038 & 0.031  \\ 
		\textit{\N{GW190521a}} & \textit{0.243} & \textit{0.158} & \textit{0.007} & \textit{0.264} & \textit{0.202} \\
		\textit{\N{GW191109}} & \textit{0.095} & \textit{0.227} & \textit{0.070} & \textit{0.422} & \textit{0.243}  \\ 
		\textit{\N{GW200129}} & \textit{0.459} & \textit{0.005} & \textit{0.330} & \textit{0.051} & \textit{0.378} \\ \hline	
		
	\end{tabular}
	\caption{$D_{JS}^{\chi_{\rm p}, \rm prior}$ values in bit computed between the posterior of $\chi_{\rm p}$ obtained with our analysis and the prior distribution conditioned to $\chi_{\rm eff}$, for the posteriors recovered with the different waveforms and the combined one. Higher values of $D_{JS}^{\chi_{\rm p}, \rm prior}$ mean a larger difference between the posterior and prior distribution. Given that the JS divergence measures the dissimilarity of two distributions by quantifying the information that would be lost if we try to describe one distribution with the other, there does not exist a general fixed threshold value above which we can claim that the two distributions are significantly different. In Ref~\cite{LIGOScientific:2020ibl}, when looking at the recovery of the $\chi_{\rm p}$ parameter, further investigations were carried out for events with $D_{JS}^{\chi_{\rm p}, \rm prior} > 0.05$. However, possible threshold values strongly depend on the context.
		Events for which we find values of $\chi_{\rm p}$ significantly different from the prior are marked in italic.}
	\label{tab:js-prior}
\end{table}
\endgroup

The detector frame masses and spins estimated with the various models are reported in Table~\ref{tab:mass-pars} and Table~\ref{tab:spin-pars}, respectively, while Table~\ref{tab:log_lik} shows the median value, along with its $90\%$ confidence interval, of the distribution of the logarithm of the samples' likelihood, $\log \mathcal{L}$. In general, we expect that higher values of $\log \mathcal{L}$ correspond to a higher probability for a given model. However, Table~\ref{tab:log_lik} reports the median of the recovered $\log \mathcal{L}$ distribution, therefore, since the shape of the distribution will affect the median value, in some cases the model with the largest $\log \mathcal{L}$ value might not correspond to the model with the largest probability.

\begin{figure*}
	\centering
	\includegraphics[width=1\textwidth]{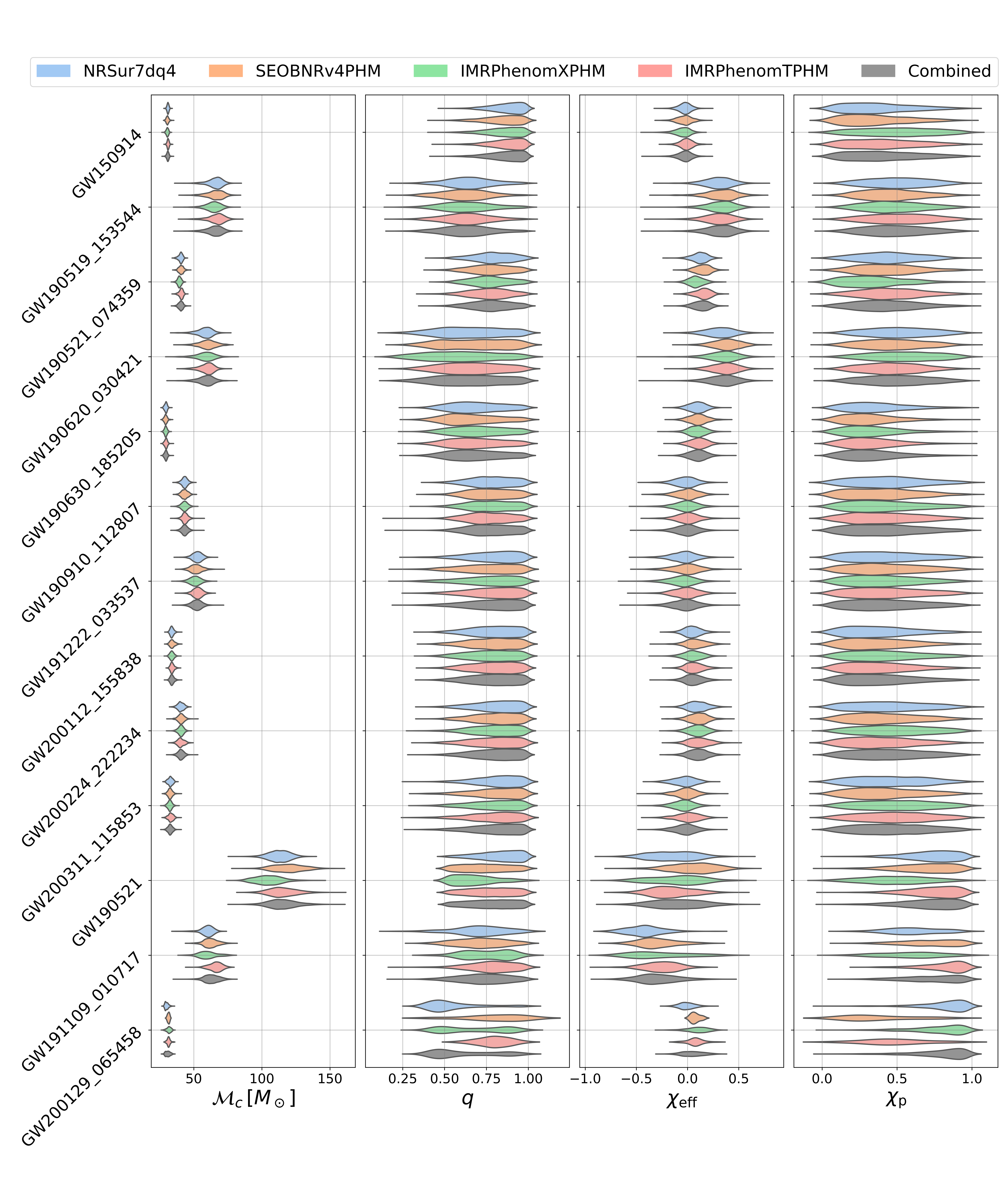}
	\caption{Posterior probability densities for $\mathcal{M}_c$, $q$, $\chi_{\rm eff}$, and $\chi_{\rm p}$ as recovered with the different waveform approximants and for the combined posterior, for all the events analyzed.}
	\label{fig:all_pe}
\end{figure*}

Regarding the spins, information is reported through the \emph{effective inspiral spin}
\begin{equation}
	\chi_{\rm eff} = \frac{\left(m_1 \chi_{1, \parallel} + m_2 \chi_{2, \parallel}\right)}{M},
\end{equation}
with $\chi_{1,\parallel}, \chi_{2,\parallel}$ being the spin components parallel to the angular momentum, and the \emph{effective precession spin}
\begin{equation}
	\chi_{\rm p}  = \max \left\{ \chi_{1, \bot}, \frac{q \left( 4q + 3 \right)}{4 + 3q} \chi_{2, \bot} \right\},
\end{equation}
where $\chi_{1,\bot}, \chi_{2,\bot}$ are the spin components perpendicular to the angular momentum.
Figure ~\ref{fig:all_pe} shows the posterior probability density of $\mathcal{M}_c$, $q$, $\chi_{\rm eff}$, and $\chi_{\rm p}$ for all the events, comparing the posteriors recovered with the different waveform models.
We can usually place only weak constraints on $\chi_{\rm p}$. Thus, its posterior distribution is heavily affected by the prior one, which in turn is determined by the source parameters $\chi_1, \chi_2$, and $q$, and peaks at non-zero values of $\chi_{\rm p}$ also in the absence of precession. Therefore, recovering a non-zero value of $\chi_{\rm p}$ does not constitute sufficient evidence of precession, but we need to check if the posterior distribution is significantly different from the prior one. This is evaluated through the Jensen-Shannon (JS) divergence \cite{Lin:1991zzm}, which estimates the difference between two probability distributions $p_1$ and $p_2$ as {\small
\begin{equation}
D_{\rm JS} = \frac{1}{2} \left[ \sum_x p_1(x) \log \left(\frac{p_1(x)}{m(x)}\right) +\sum_x p_2(x) \log \left(\frac{p_2(x)}{m(x)}\right)  \right], 
\end{equation}
}with $m(x) = 0.5 (p_1(x) + p_2(x))$. Table~\ref{tab:js-prior} shows the JS divergence values for $\chi_{\rm p}$ posteriors with respect to their prior distribution, $D_{JS}^{\chi_{\rm p}, \rm prior}$. We also compare our results with the ones from LVK analyses in Table~\ref{tab:js-pe}, where the difference between the posterior distributions is again evaluated as a JS divergence.
Furthermore, the probabilities recovered for each model, together with their errors, are reported in Table~\ref{tab:probs} for all the events analyzed.


\renewcommand*{\arraystretch}{2}
\begin{table*}
	\begin{tabular}{c|c c c c|c c c c }
		\hline
		&\multicolumn{4}{c}{\centering{\xphm}} & \multicolumn{4}{c}{\centering{\seob}} \\ \hline
		Event & $D_{JS}^{\chi_{\rm p}}$ & $D_{JS}^{\chi_{\rm eff}}$ & $D_{JS}^{\mathcal{M}_c}$ & $D_{JS}^{q}$ & $D_{JS}^{\chi_{\rm p}}$ & $D_{JS}^{\chi_{\rm eff}}$ & $D_{JS}^{\mathcal{M}_c}$ & $D_{JS}^{q}$ \\ \hline
		\N{GW150914} & 0.006 & 0.001 & 0.007 & 0.005 & 0.005 & 0.076 & 0.032 & 0.052 \\
		\N{GW190519} & 0.002 & 0.001 & 0.001 & 0.001 & 0.011 & 0.002 & 0.014 & 0.007 \\
		\N{GW190521b} & 0.007 & 0.004 & 0.005 & 0.001 & 0.024 & 0.033 & 0.005 & 0.007 \\
		\N{GW190620} & 0.005 & 0.001 & 0.007 & 0.001 & 0.001 & 0.001 & 0.006 & 0.001 \\
		\N{GW190630} & 0.002 & 0.001 & 0.004 & 0.002 & 0.010 & 0.022 & 0.025 & 0.015 \\
		\N{GW190910} & 0.002 & 0.000 & 0.002 & 0.001 & 0.007 & 0.002 & 0.015 & 0.009 \\
		\N{GW200112} & 0.002 & 0.007 & 0.004 & 0.003 & 0.009 & 0.018 & 0.014 & 0.026 \\
		\N{GW200224} & 0.001 & 0.003 & 0.006 & 0.001 & 0.004 & 0.016 & 0.018 & 0.006 \\
		\N{GW200311} & 0.002 & 0.001 & 0.001 & 0.001 & 0.006 & 0.006 & 0.015 & 0.014 \\ 
		\N{GW190521a} & 0.019 & 0.003 & 0.066 & 0.075 & 0.020 & 0.003 & 0.018 & 0.035 \\
		\N{GW191109} & 0.024 & 0.006 & 0.012 & 0.006 & 0.029 & 0.011 & 0.016 & 0.007  \\
		\N{GW200129} & 0.003 & 0.010 & 0.005 & 0.008 & 0.139 & 0.046 & 0.137 & 0.141 \\ \hline

	\end{tabular}
	\caption{Values of Jensen-Shannon divergence for $\chi_{\rm p}$, $\chi_{\rm eff}$, $\mathcal{M}_c$, and $q$, computed between the posteriors recovered by our analysis and the LVK ones \cite{LIGOScientific:2021usb,LIGOScientific:2021djp} for the available waveforms, \xphm and \seob. As mentioned for Table~\ref{tab:js-pe}, lower values of the JS divergence correspond to more similar distributions.}
	\label{tab:js-pe}
\end{table*}

\subsection{Single events}

In this section, we comment on the individual event recoveries with the different waveform models.

\subsubsection{\N{GW150914}}

For this event, the parameters and the log-likelihoods (see Table~\ref{tab:log_lik}) recovered are consistent for all four models. The recovered values for the source parameters can be found in Table~\ref{tab:mass-pars}-\ref{tab:spin-pars}, and are consistent with the LVK results in~\cite{LIGOScientific:2021usb,LIGOScientific:2018mvr}, as shown in Table~\ref{tab:js-pe}.
The probabilities for each approximant are reported in Table~\ref{tab:probs}, where we see a slight preference for the \tphm model. 

\subsubsection{\N{GW190519}}

In this case, data show a preference for \xphm (see Table~\ref{tab:probs}), although parameter estimates and log-likelihood values are consistent for all the models (see Tables~\ref{tab:mass-pars},~\ref{tab:spin-pars} and~\ref{tab:log_lik}). 
We find support for positive, non-zero values of $\chi_{\rm eff}$. This is consistent with the results reported in Ref.~\cite{LIGOScientific:2021usb}.

\subsubsection{\N{GW190521b}}

This event shows a preference for the \seob approximant (cf. Table~\ref{tab:probs}), although the recovered parameters and log-likelihood values, reported in Tables~\ref{tab:mass-pars},~\ref{tab:spin-pars} and~\ref{tab:log_lik}, respectively, are similar for all four models. Also in this case, our results are consistent with the ones in the LVK papers~\cite{LIGOScientific:2021usb} see Table~\ref{tab:js-pe}, and we find no evidence of precession. 

\subsubsection{\N{GW190620}}
The LVK studies report this source as a BBH binary with high effective spin $\chi_{\rm eff}$. In our re-analysis, we find all the waveform families to perform comparably and return consistent estimates of parameters as well as the values of log-likelihood (see Tables~\ref{tab:mass-pars},~\ref{tab:spin-pars} and~\ref{tab:log_lik}, respectively). Moreover, the existing LVK analyses on this event with \xphm and \seob return consistent results with ours, as shown in Table~\ref{tab:js-pe}). We also find support for positive values of $\chi_{\rm eff}$. The estimates of intrinsic parameters from different models are consistent with each other, however, from the values of posterior probability (see Table~\ref{tab:probs}), \nrsur seems to be the most favored.

\subsubsection{\N{GW190630}}
We find consistent estimates of parameters and log-likelihoods among all models compared (cf. Tables~\ref{tab:mass-pars},~\ref{tab:spin-pars} and~\ref{tab:log_lik}), and no evidence for spin. Among the four models considered, \nrsur and \tphm seem to be most preferred by the data, with almost the same probability see Table~\ref{tab:probs}. 

\subsubsection{\N{GW190910}}
This event again returns very consistent estimates of log-likelihoods and intrinsic parameters among the different models (see. Tables~\ref{tab:mass-pars},~\ref{tab:spin-pars} and~\ref{tab:log_lik}). In particular, we find no evidence for spins. From the values of posterior probabilities supported by all waveforms, we also note that the data have an almost equal preference for all models (cf. Table~\ref{tab:probs}). 

\subsubsection{\N{GW191222}}
Although the returned parameter estimates, as well as log-likelihood values, are quite similar (see. Tables~\ref{tab:mass-pars},~\ref{tab:spin-pars} and~\ref{tab:log_lik}), \tphm seems to be the most favored model (see Table~\ref{tab:probs}), while the least favored model is \xphm. We find no evidence for spins. 

\begin{figure}
	\centering
	\includegraphics[width=0.5\textwidth,height=0.6\textwidth]{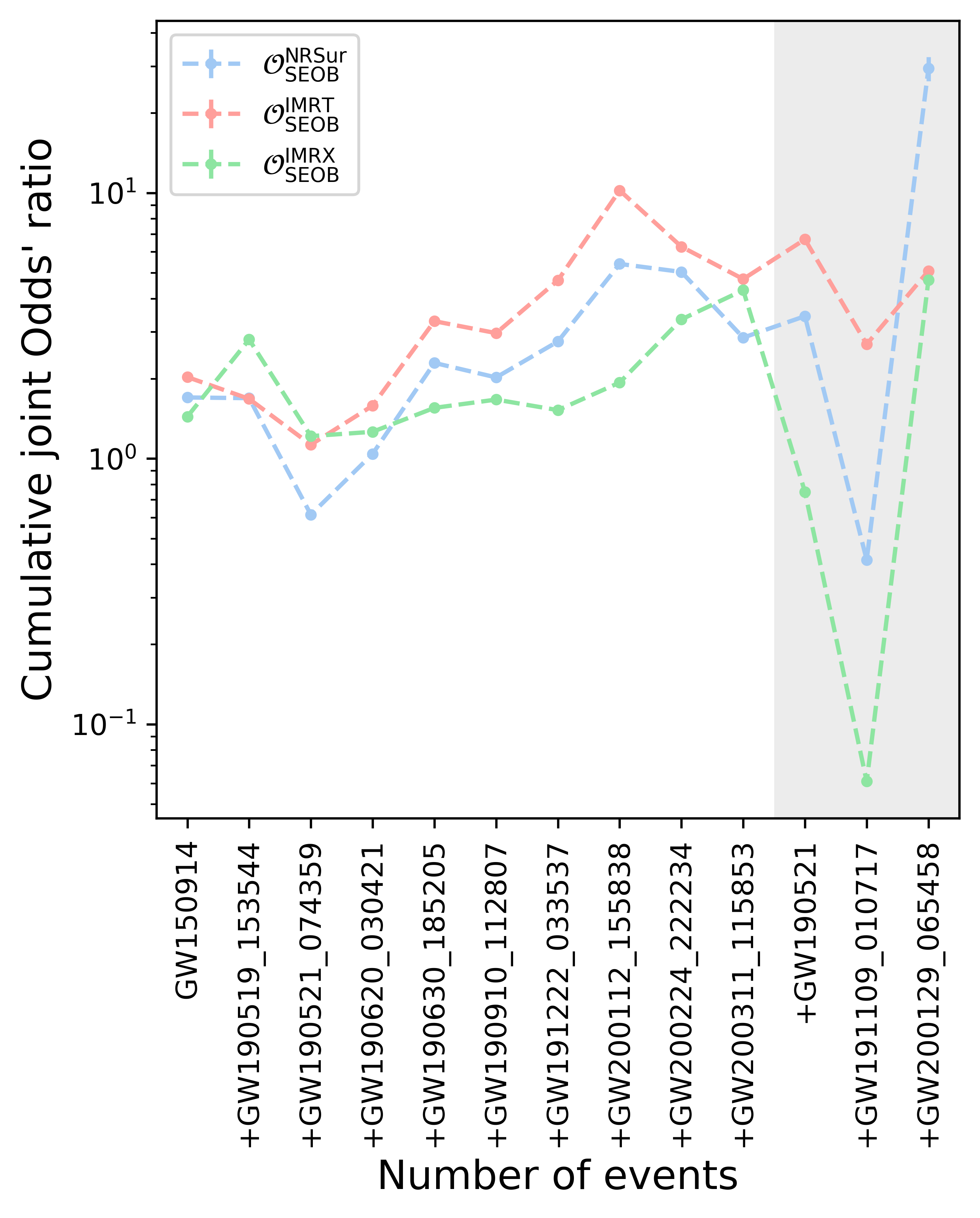}
	\caption{Evolution of the joint odds ratio for each approximant, with respect to \seob, as events are added; for any one event shown on the x-axis, the joint odds ratio is calculated from all the events occurring to the left of that event. The events in the gray-shaded area are affected by possible data quality issues. 
		Note that the symmetric error bars $1\sigma$ are included in the data points but too small to be discernible.}
	\label{fig:joint_odds}
\end{figure}

\begingroup
\renewcommand*{\arraystretch}{2}
\begin{table*}
	\begin{tabular}{c|c|c|c|c}
		\hline
		Event & NRSur & SEOB & IMRX & IMRT \\ \hline
		\N{GW150914} & 27.55 $\pm$ 0.7 & 16.22 $\pm$ 0.8 & 23.34 $\pm$ 0.7 & 32.88 $\pm$ 0.7 \\
		\N{GW190519} & 20.82 $\pm$ 0.6 & 20.95 $\pm$ 0.6 & 40.87 $\pm$ 0.5 & 17.35 $\pm$ 0.6 \\
		\N{GW190521b} & 14.76 $\pm$ 1.2 & 40.50 $\pm$ 1.0 & 17.53 $\pm$ 1.2 & 27.22 $\pm$ 1.1 \\
		\N{GW190620} & 32.98 $\pm$ 0.6 & 19.48 $\pm$ 0.6 & 20.22 $\pm$ 0.6 & 27.32 $\pm$ 0.6 \\
		\N{GW190630} & 33.79 $\pm$ 0.6 & 15.36 $\pm$ 0.6 & 18.90 $\pm$ 0.6 & 31.95 $\pm$ 0.6 \\
		\N{GW190910} & 22.86 $\pm$ 0.6 & 25.92 $\pm$ 0.6 & 27.85 $\pm$ 0.6 & 23.37 $\pm$ 0.6 \\
		\N{GW191222} & 28.11 $\pm$ 0.5 & 20.58 $\pm$ 0.6 & 18.78 $\pm$ 0.6 & 32.53 $\pm$ 0.5 \\
		\N{GW200112} & 30.56 $\pm$ 0.6 & 15.61 $\pm$ 0.6 & 19.82 $\pm$ 0.6 & 34.01 $\pm$ 0.5 \\
		\N{GW200224} & 21.82 $\pm$ 0.6 & 23.39 $\pm$ 0.6 & 40.43 $\pm$ 0.5 & 14.36 $\pm$ 0.7 \\
		\N{GW200311} & 15.68 $\pm$ 0.6 & 27.70 $\pm$ 0.6 & 35.69 $\pm$ 0.6 & 20.93 $\pm$ 0.6 \\
		\textit{\N{GW190521a}} & \textit{31.78 $\pm$ 0.6} & \textit{26.39 $\pm$ 0.6} & \textit{4.60 $\pm$ 0.7} & \textit{37.23 $\pm$ 0.5} \\
		\textit{\N{GW191109}} & \textit{7.54 $\pm$ 1.6} & \textit{62.29 $\pm$ 1.0} & \textit{5.06 $\pm$ 1.7} & \textit{25.11 $\pm$ 1.5} \\
		\textit{\N{GW200129}} & \textit{46.94 $\pm$ 1.4} & \textit{0.66}$^{\mathit{+1.9}}_{\mathit{-0.66}}$ & \textit{51.14 $\pm$ 1.3} & \textit{1.25} $^{\mathit{+1.9}}_{\mathit{-1.25}}$ \\ \hline
		
	\end{tabular}
	\caption{Probability percentages, including errors, for each model in the different events. Events that strongly favor or disfavor some of the models are marked in italic.}
	\label{tab:probs}
\end{table*}
\endgroup

\subsubsection{\N{GW200112}}

We recover similar probabilities for all the approximants, with \seob slightly disfavored and \tphm slightly favored, as shown in Table~\ref{tab:probs}. Consistently, we find no significant difference between the recovered parameters and log-likelihoods for the different waveforms (see. Tables~\ref{tab:mass-pars},~\ref{tab:spin-pars} and~\ref{tab:log_lik}). 
The \xphm and \seob posteriors estimated by our study are consistent with the LVK ones, as reported in Table~\ref{tab:js-pe}.

\subsubsection{\N{GW200224}}

For this event the recovered parameters and log-likelihood values are consistent for the different waveforms (see. Tables~\ref{tab:mass-pars},~\ref{tab:spin-pars} and~\ref{tab:log_lik}). We find a slight preference for \xphm, cf.~Table~\ref{tab:probs}. Our results for both \xphm and \seob are consistent with the LVK ones, as shown in Table~\ref{tab:js-pe}. We do not find support for precession. 

\subsubsection{\N{GW200311}}
Specific to this event, we find no evidence of spin and consistent source parameters and log-likelihood estimates among all models, as reported in Tables~\ref{tab:mass-pars},~\ref{tab:spin-pars} and~\ref{tab:log_lik}. However, \xphm seems to be the most favored approximant by the event (cf. Table~\ref{tab:probs}).

\subsubsection{\N{GW190521a}}

\N{GW190521a} is the most massive event detected so far, and one among the ones with the strongest signature of higher-order modes in the signal~\cite{LIGOScientific:2020iuh,LIGOScientific:2020ufj}. The consequently high values needed for the prior on chirp mass, combined with the employed prior on mass ratio, cause potential issues with the \tphm model since the computed peak frequency for the $\ell,m = (2,2)$ mode might be below the 20~Hz low-frequency cutoff used for our analysis. To avoid this issue, for this event, we adjust the prior on mass ratio such that $q \le 2$. The recovered values for mass and spin parameters are reported in Table~\ref{tab:mass-pars} and Table~\ref{tab:spin-pars}, respectively. They are consistent with the results in Ref.~\cite{LIGOScientific:2021usb} (cf. Table~\ref{tab:js-pe}), where, however, only the \xphm and \seob approximants were used\footnote{In Ref.~\cite{LIGOScientific:2021usb}, further analyses computed the precession SNR to be too small to claim the presence of strong evidence for precession.}, and with the \nrsur results first shown in the discovery paper~\cite{LIGOScientific:2020iuh}. We find evidence of precession for the \nrsur, \seob, and \tphm models, cf. Table~\ref{tab:js-prior}. The probabilities for the different approximants are shown in Table~\ref{tab:probs}: the \tphm model is slightly favored over the other ones, while \xphm is strongly disfavored. Interestingly, these findings are consistent with the fact that the \xphm model provides a less accurate description of precession in the ringdown phase: being a frequency-domain model, it is not straightforward to compute a specific closed-form ansatz for the Euler angles during the ringdown, and therefore the same prescription for the inspiral is employed; moreover, the stationary phase approximation is used in the whole waveform, although it is not adequate for the merger and ringdown. These limitations become more evident in the case of signals where the merger and ringdown phase prevail, like \N{GW190521a}. 
Nevertheless, the extremely short duration of this event and the lack of the inspiral part of the signal make it difficult to draw clear conclusions. \\
Many works investigated this event from different perspectives and explored the possible processes that lead to the formation of such a system. One of the most investigated hypotheses is the presence of eccentricity~\cite{Romero-Shaw:2020thy, Gayathri:2020coq, Gayathri:2020mra}, which could mimic precession~\cite{Xu:2022zza,CalderonBustillo:2020xms}.
Multiple alternative scenarios that could lead to the emission of this signal have been proposed, like dynamical capture in hyperbolic orbits~\cite{Gamba:2021gap}, a primordial BH merger~\cite{DeLuca:2020sae}, and a high-mass BH-disk system~\cite{Shibata:2021sau}. In Ref.~\cite{Fishbach:2020qag}, an analysis of this event with a population-based prior led to the conclusion that neither of the component masses lies in the pair-instability supernova mass gap. In Ref.~\cite{Nitz:2020mga}, the use of a high-mass prior showed the possibility of \N{GW190521a} being an intermediate-mass-ratio BBH merger. However, a further investigation carried out in Ref.~\cite{Estelles:2021jnz}, where different precession prescriptions and higher-order-mode contents were investigated with the \xphm and \tphm models, showed that, despite the presence of a multimodal likelihood for the mass ratio parameter, the peaks are characterized by very different probabilities. The parameters recovered by our analysis are consistent with both the \xphm and \tphm results in Ref.~\cite{Estelles:2021jnz}, when using models with the same settings.

\subsubsection{\N{GW191109}}

We find \seob to be the most favored model, as reported in Table~\ref{tab:probs}. We also recover a high probability for \tphm, while \nrsur and \xphm are strongly disfavored. We find evidence of non-zero $\chi_{\rm p}$ with both \seob and \tphm, but not for the other two models, as shown in Table~\ref{tab:spin-pars} and \ref{tab:js-prior}. 
For all models, we find significant support for negative values of $\chi_{\rm eff}$ (cf. Table~\ref{tab:spin-pars}), confirming the results in Refs.~\cite{LIGOScientific:2021djp,Zhang:2023fpp}. In the latter, the possibility of formation by dynamical capture for the binary generating this event is discussed. However, \N{GW191109} was among the O3 events that required data mitigation due to the presence of glitches. In particular, \N{GW191109} was affected by a glitch in both the detectors online at the time of the event, in the frequency range 25-45~Hz for Hanford and 20-32~Hz for Livingston. As shown in Ref.~\cite{Davis:2022ird}, different deglitching procedures influence the posteriors obtained for both $\chi_{\rm eff}$ and $\chi_{\rm p}$. In particular, if the Livingston data are analyzed only for frequencies larger than 40~Hz, the support for negative $\chi_{\rm eff}$ disappears. However, this result is not sufficient to label the negative support of $\chi_{\rm eff}$ as a noise artifact, since most of the spin information comes from low frequencies, and, being \N{GW191109} already a signal with a short inspiral, removing the low-frequency part discards most of the information, yielding non-informative results. The presence of glitches overlapping a significant part of the inspiral for both the detectors is also regarded as the most likely cause for deviations from general relativity found for this event by some LVK pipelines~\cite{LIGOScientific:2021sio}.

\subsubsection{\N{GW200129}}

We find a strong preference for \nrsur and \xphm, while the probability for \seob and \tphm is close to zero (cf. Table~\ref{tab:probs}). This discrepancy is reflected in the posteriors of $\chi_{\rm p}$, with \nrsur and \xphm finding strong evidence for high $\chi_{\rm p}$ values, cf.~Table~\ref{tab:js-prior}, while for the other two models results are dominated by the prior. This is consistent with what was found in the LVK GWTC-3 analysis \cite{LIGOScientific:2021djp}, where \xphm recovers $\chi_{\rm p}$ and \seob does not. 
In Ref.~\cite{Hannam:2021pit}, strong evidence for precession was found when analyzing this event with the \nrsur model. 
For this event, precession was measured also in Ref.~\cite{Varma:2022pld}, where the recoil velocity was also estimated.
The main difference between these two works and the LVK analysis \cite{LIGOScientific:2021djp}, which did not find conclusive evidence of precession, is that in the latter data were analyzed only with the \xphm and \seob approximants.
In Refs.~\cite{Hannam:2021pit} and \cite{Varma:2022pld}, the \nrsur model was used, because, being generated from NR simulations, it is expected to be more accurate, as shown by the mismatch computation in Ref.~\cite{Hannam:2021pit}. However, in our study, we do not find an overall preference for \nrsur. 
GW200129 data were affected by a glitch overlapping the event in the Livingston detector~\cite{Davis:2022ird},
therefore, in our analysis, we used the deglitched data, as was done in Ref.~\cite{LIGOScientific:2021djp}. Reference~\cite{Payne:2022spz} explores the influence of data quality issues for this event, finding that the evidence for precession comes exclusively from the Livingston strain of data between 20-50 Hz, where such issues are present. 

\begingroup
\renewcommand*{\arraystretch}{2}
\begin{table*}
	\begin{tabular}{c | c c c c c c}
		\hline
		& $\mathcal{O}^\mathrm{NRSur}_\mathrm{SEOB}$ & $\mathcal{O}^\mathrm{IMRX}_\mathrm{SEOB}$ & $\mathcal{O}^\mathrm{IMRT}_\mathrm{SEOB}$ & $\mathcal{O}^\mathrm{NRSur}_\mathrm{IMRX}$ & $\mathcal{O}^\mathrm{NRSur}_\mathrm{IMRT}$ & $\mathcal{O}^\mathrm{IMRX}_\mathrm{IMRT}$\\ \hline
		All events & $29.43 \pm 1.11$ & $4.70 \pm 0.07$ & $5.09 \pm 0.08$ & $6.26 \pm 0.11$ & $5.78 \pm 0.10$ & $0.92 \pm 0.01$ \\ \hline
		No \N{GW200129} & $0.42 \pm 0.00$ & $0.06 \pm 0.00$ & $2.69 \pm 0.03$ & $6.82 \pm 0.12$ & $0.15 \pm 0.00$ & $0.02 \pm 0.00$ \\ 
		No \N{GW190521a}  & $24.44 \pm 0.84$ & $26.99 \pm 0.97$ & $3.61 \pm 0.05$ & $0.91 \pm 0.01$ & $6.77 \pm 0.12$ & $7.48 \pm 0.14$ \\
		No \N{GW191109} & $243.31 \pm 26.35$ & $57.84 \pm 3.05$ & $12.62 \pm 0.31$ & $4.21 \pm 0.06$ & $19.27 \pm 0.59$ & $4.58 \pm 0.07$ \\
		Without all three & $2.85 \pm 0.03$ & $4.30 \pm 0.06$ & $4.74 \pm 0.07$ & $0.66 \pm 0.00$ & $0.60 \pm 0.00$ & $0.91 \pm 0.01$ \\ \hline
	\end{tabular}
	\caption{Joint odds ratios including errors. We report results for all the events combined and results without the events that show a strong preference for some models.}
	\label{tab:joint-odds}
\end{table*}
\endgroup

\subsection{Combined events}

Figure~\ref{fig:joint_odds} shows the cumulative joint odds ratio as a function of the number of events, while Table~\ref{tab:joint-odds} reports the odds ratio values obtained by combining information from all the sources analyzed. We do not find a specific approximant being preferred or disfavored consistently for all the events. Combining results for all the 13 sources, the \nrsur model results favored with respect to \seob, with an odds ratio of 29.43. However, this value is dominated by the results for \N{GW200129}, and without this event the odds ratio becomes 0.46. This is unexpected, because \nrsur, being fully informed by NR simulations, is assumed to be the most accurate model and therefore to describe the data best. Table~\ref{tab:joint-odds} shows also how odds ratios change with the three events with a strong preference for one of the models: while \N{GW200129} is responsible for \nrsur being favored over \seob, \N{GW191109}, which instead finds a significant preference for \seob and \tphm, balances this result; if we do not take \N{GW191109} into account, \nrsur and \xphm are strongly favored over \seob, with an odds ratio of 243.31 and 57.84 respectively. In addition, without this event, $\mathcal{O}^{\rm IMRT}_{\rm SEOB} = 12.62$, and $\mathcal{O}^{\rm NRSur}_{\rm IMRT} = 19.27$. Similarly, the results from \N{GW190521a} heavily influence the final odds ratio for \xphm: if we do not include this event, we obtain $\mathcal{O}^\mathrm{IMRX}_\mathrm{SEOB} = 26.99$. 
Without these three sources, we find no significant preference for any of the models.

\begin{figure*}
	\centering
	\includegraphics[width=1\textwidth]{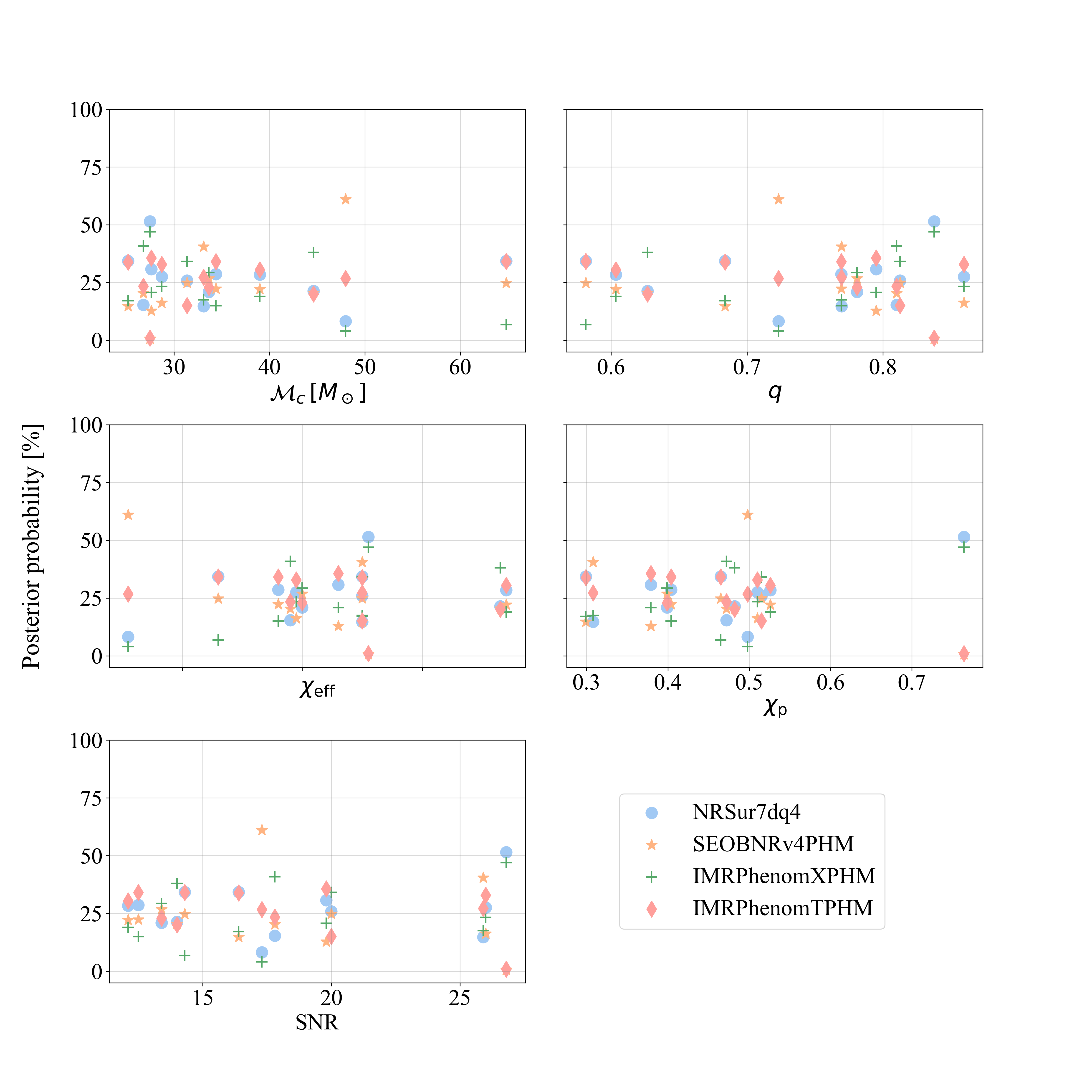}
	\caption{Posterior probability for the different approximants as a function of the LVK estimated values of $\mathcal{M}_c$ (top left panel), $q$ (top right), $\chi_{\rm eff}$ (middle left), $\chi_{\rm p}$ (middle right), and SNR (bottom left panel).}
	\label{fig:params_trends}
\end{figure*}

We look for possible trends for the preference of given approximants with respect to the binary parameters, which would point to the waveforms with the best description for specific regions of the parameter space. Figure~\ref{fig:params_trends} shows the probabilities recovered for the different models as a function of the source's mass and spin parameters, and the network optimal matched-filter SNR, as computed by the parameter estimation analyses in the catalog papers~\cite{LIGOScientific:2021usb,LIGOScientific:2021djp}. We do not find any trends with respect to the binary parameters or the signal SNR.

Interestingly, we find that for all the events that show a strong preference for one of the models, i.e., \N{GW190521a}, \N{GW191109}, and \N{GW200129}, the preferred models are not the same, but in each case are the ones that recover precession. This is particularly evident in the case of \N{GW190521a}, where \xphm does not recover evidence of precession and has a probability only of roughly $4\%$, while the other models, which show evidence supporting non-zero values of $\chi_{\rm p}$, have all a probability $\sim 30 \%$. Although, as mentioned, the results for these events might be biased by their short duration or potential data quality issues, the fact that a given model recovers precession better than another one systematically implies a higher probability. Evidence for this behavior is supported by the fact that the preferred models are different for the three events, leaving the recovery of precession as the only element systematically connected to higher probability values.

\section{Summary}
\label{sec:summary}

We analyzed the 13 events with the highest mass and moderate to high SNR among the ones detected so far by Advanced LIGO and Advanced Virgo, using the ``hypermodels'' technique developed in Ref.~\cite{Ashton:2021cub}. This method allows us to sample directly over different waveform approximants, in order to determine which one is favored by the data. We analyzed data with four different approximants, all including precession and higher-order modes: \nrsur, \seob, \xphm, and \tphm. For each event, we recover the source parameters, finding both mass and spin parameters to be in agreement with the LVK results, cf.~Table~\ref{tab:js-pe}. For three events, \N{GW191109}, \N{GW200129}, and \N{GW190521a}, we recover non-zero values for the effective precession spin parameter, with a distribution significantly different from the prior one. These events are also the ones for which we find a strong preference for some models over the other ones, although the preferred approximants are different. \N{GW191109} shows a strong preference for \seob, with \nrsur and \xphm being disfavored. On the other hand, for \N{GW200129}, \nrsur and \xphm are strongly favored, and the probability for \seob and \tphm is close to zero. Finally, \N{GW190521a} recovers a very low probability, roughly $4\%$, for \xphm, while the other models do not show significant differences among them. However, \N{GW191109} and \N{GW200129} data were affected by glitches~\cite{Davis:2022ird}, and the short duration of \N{GW190521a} implies that we could not see its inspiral phase; therefore, we cannot draw clear conclusions about these events. Nonetheless, we systematically find that the models recovering evidence for non-zero values of $\chi_{\rm p}$ are the ones with the higher probabilities. 
For all the other events, we recover only slight preferences for a given approximant, with the recovered parameters' posteriors and log-likelihoods being similar.
Overall, we do not find one model to be consistently preferred over the others. This is unexpected, considering that we included \nrsur in the analysis, which is predicted to be the most accurate model for high-mass signals, being interpolated from NR simulations.
The odds ratios combined over all the sources show \nrsur being favored over \seob, with  $\mathcal{O}^\mathrm{NRSur}_\mathrm{SEOB} = 29.43 $, while for \xphm and \tphm we find $\mathcal{O}^\mathrm{IMRX}_\mathrm{SEOB} = 4.70$ and $\mathcal{O}^\mathrm{IMRT}_\mathrm{SEOB} = 5.09$ respectively.
However, this result is mostly determined by \N{GW200129}, for which \seob and \tphm probabilities are close to zero. If we remove this event from the combined odds ratio calculation, we obtain $\mathcal{O}^\mathrm{NRSur}_\mathrm{SEOB} = 0.42 $. 
Finally, if we do not take into account the three sources favoring one of the approximants, we find no significant preference for any of the models.


\begin{acknowledgments} 

We thank Vijay Varma for his help with the \nrsur model. A.P. and C.V.D.B.~are supported by the research programme of the Netherlands Organisation for Scientific Research (NWO). A.S. thanks the Alexander von Humboldt foundation in Germany for a
Humboldt fellowship for postdoctoral researchers.
This work was co-funded by the European Union (ERC, SMArt, project number 101076369). Views and opinions expressed are however those of the author(s) only and do not necessarily reflect those of the European Union or the European Research Council. Neither the European Union nor the granting authority can be held responsible for them.

We are grateful for computational resources provided by Cardiff University, and
funded by an STFC grant (ST/I006285/1) supporting UK Involvement in the Operation of Advanced
LIGO. 
This research has made use of data or software obtained from the Gravitational Wave Open Science Center (gwosc.org), a service of the LIGO Scientific Collaboration, the Virgo Collaboration, and KAGRA. This material is based upon work supported by NSF's LIGO Laboratory which is a major facility fully funded by the National Science Foundation, as well as the Science and Technology Facilities Council (STFC) of the United Kingdom, the Max-Planck-Society (MPS), and the State of Niedersachsen/Germany for support of the construction of Advanced LIGO and construction and operation of the GEO600 detector. Additional support for Advanced LIGO was provided by the Australian Research Council. Virgo is funded, through the European Gravitational Observatory (EGO), by the French Centre National de Recherche Scientifique (CNRS), the Italian Istituto Nazionale di Fisica Nucleare (INFN) and the Dutch Nikhef, with contributions by institutions from Belgium, Germany, Greece, Hungary, Ireland, Japan, Monaco, Poland, Portugal, Spain. KAGRA is supported by Ministry of Education, Culture, Sports, Science and Technology (MEXT), Japan Society for the Promotion of Science (JSPS) in Japan; National Research Foundation (NRF) and Ministry of Science and ICT (MSIT) in Korea; Academia Sinica (AS) and National Science and Technology Council (NSTC) in Taiwan.

\end{acknowledgments}

\appendix

\begin{figure*}
	\centering
	\includegraphics[width=1\textwidth]{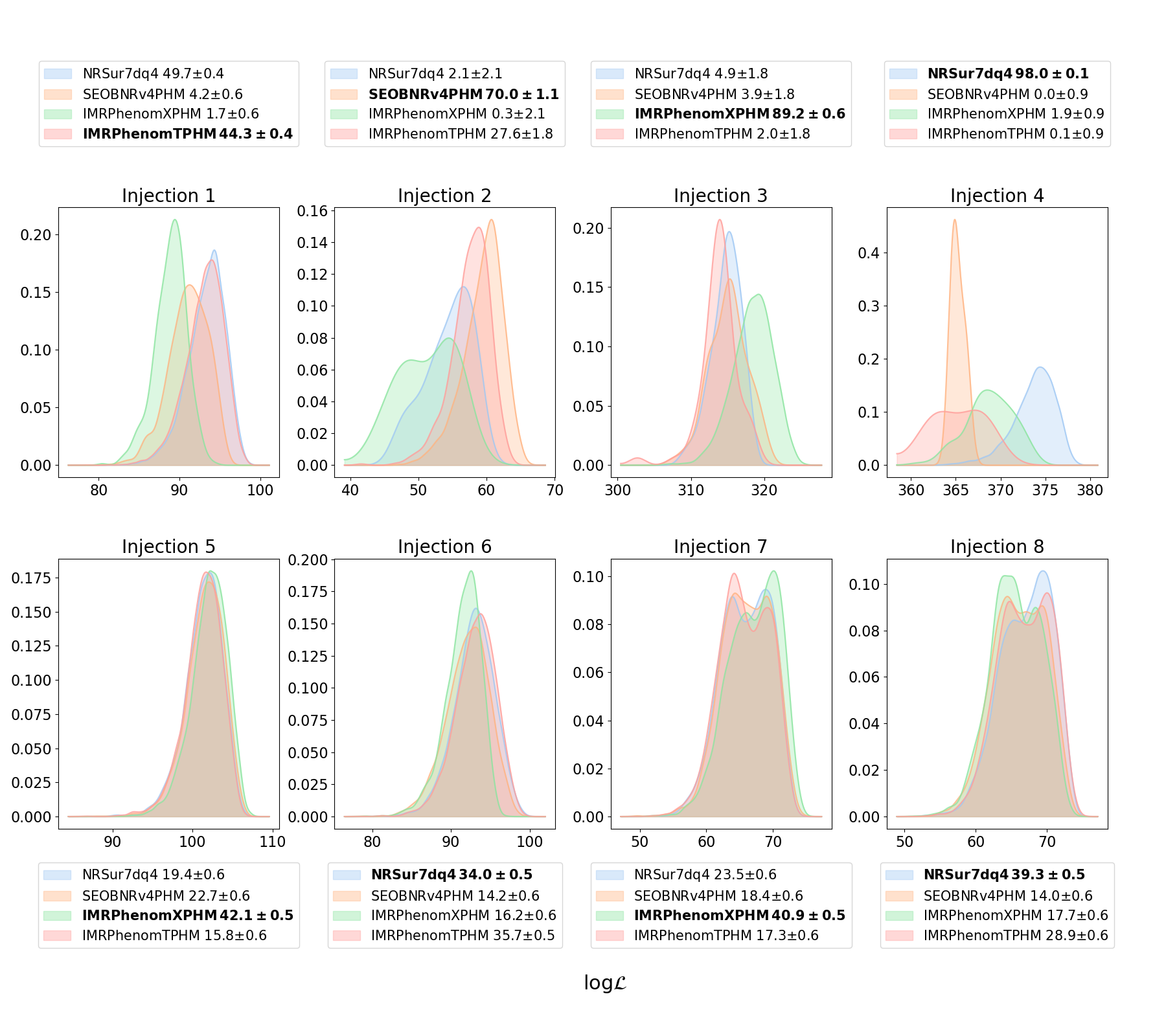}
	\caption{Probability density distributions for $\log \mathcal{L}$ for the different models considered in the analysis. The legend reports the recovered probability percentages, including error; the model marked in bold is the one used for the injection.}
	\label{fig:injections}
\end{figure*}

\renewcommand*{\arraystretch}{2}
\begin{table*}
	\begin{tabular}{c | c | >{\centering\arraybackslash}p{1.5cm} >{\centering\arraybackslash}p{1.5cm} >{\centering\arraybackslash}p{1.5cm} >{\centering\arraybackslash}p{1.5cm} >{\centering\arraybackslash}p{1.5cm} >{\centering\arraybackslash}p{1.5cm} >{\centering\arraybackslash}p{1.5cm}}
		\hline
		& Model & $\mathcal{M}_c$ [$\, M_\odot $] & $q$ & $a_1$ & $\theta_1$ [rad] & $a_2$ & $\theta_2$ [rad] & $D_L$ [Mpc] \\ \hline
		Injection 1 & \tphm & 108.79 & 0.92 & 0.97 & 2.59 & 0.93 & 1.66 & 2751.72 \\
		Injection 2 & \seob & 71.32 & 0.54 & 0.99 & 1.12 & 0.81 & 1.96 & 3488.44 \\
		Injection 3 & \xphm & 28.94 & 0.42 & 0.88 & 1.55 & 0.73 & 1.95 & 1358.51 \\
		Injection 4 & \nrsur & 28.94 & 0.42 & 0.88 & 1.55 & 0.73 & 1.95 & 1358.51 \\
		Injection 5 & \xphm & 65.72 & 0.63 & 0.81 & 1.74 & 0.68 & 1.72 & 2000.0 \\
		Injection 6 & \nrsur & 65.72 & 0.63 & 0.81 & 1.74 & 0.68 & 1.72 & 2000.0 \\
		Injection 7 & \xphm & 65.72 &  0.63 & 0.64 & 0.0 & 0.58 & 0.0 & 2000.0 \\
		Injection 8 & \nrsur & 65.72 &  0.63 & 0.64 & 0.0 & 0.58 & 0.0 & 2000.0 \\ \hline
\end{tabular}
\caption{Approximant model and parameters used for injections; $a_{1,2}$ and $\theta_{1,2}$ represent the magnitude and tilt angle of the components' spins, while $D_L$ is the luminosity distance.}
\label{tab:inj}
\end{table*}

\section{Injection runs}
\label{sec:appendix}

In addition to the analyis of real GW events, we want to prove in the following the validity of the method through an injection study. For this purpose, we perform a hypermodels analysis with the same waveform approximants and settings previously described, to analyze simulated signals in zero noise.
The details of the injections are given in Table~\ref{tab:inj}. Injection 1 and injection 2 are produced using the maximum-likelihood parameters and approximants recovered from the analyses of \N{GW190521a} and \N{GW191109}, respectively. Injection 3 and 4 are generated with the maximum-likelihood parameters of \N{GW200129} using \xphm and \nrsur, which are the models with the highest recovered probability and likelihood, respectively. For the other injections we employed the maximum-likelihood mass values recovered for \N{GW190519}, a fixed luminosity distance, and two different values of spin magnitudes and inclinations, considering injections both with \xphm and \nrsur. Figure~\ref{fig:injections} shows the probability density distributions of the recovered log-likelihoods for the different models, together with their percentage probabilities, including errors. In most cases we clearly recover the highest probability for the injected model. 
When the most favored model is not the injected one, however, the probability of the injected model is very close to the highest one. This is likely due to the fact that the two waveform descriptions are very similar, and the injected model is guaranteed to provide the best fit only at the injection point. 
To further understand why the injected model in some cases is not the most favored one, a detailed analysis of different ingredients for all employed waveform models would be required, which is however outside the scope of this paper. 
From the statistical point of view, the injection study indicates that our uncertainty on the odds might not  measure the full uncertainty. A validation of the uncertainty estimates would need multiple runs on the same data set.\\
We also note that, in order to validate the method, we performed these analyses in zero noise: in real-events analysis, the presence of noise and noise fluctuations will affect the differences between the evidences.

\bibliography{refs}

\end{document}